\documentstyle[aps,pre,preprint,epsf]{revtex}
\draft
\begin{document}

\title{Multi-step Bose-Einstein Condensation of Trapped Ideal Bose Gases}
\author{Wenji Deng
\thanks{E-mail: phwjdeng@scut.edu.cn} }
\address{CCAST (World Laboratory), Beijing 100080, China\\
and \\
Department of Physics, South China 
University of Technology, Guangzhou 510641, China
\thanks{mailing address}}

\maketitle
\vspace*{1.0cm} 

{\bf Abstract:}\ \ \ \ The Phenomenon of multi-step Bose-Einstein condensation (BEC) of a finite 
number of non-interacting bosons in  anisotropic traps has been  demonstrated 
by studing the populations on eight subsets of states.
The cusp in the specific heat is found to be associated with the
crossover between subsets of states involving Bose functions $g_{n}(z)$
of different classes, as specified by their behaviour at $z=1$.  


\newpage
\noindent {\bf 1. Introduction}

The experimental realizations \cite{exp1} of Bose-Einstein
condensation (BEC) \cite{einstein1} in trapped dilute alkali
atomic vapours have attracted much attention recently. 
Standard textbooks \cite{huang1} on statistical mechanics typically
treats the topics of 
non-interacting free Bose gas in one, two, and three dimensional boxes 
in the thermodynamic limit.  
BEC is defined as  the onset of a macroscopic occupation of the 
ground state at a finite temperature, and it is usually 
accompanied by a 
cusp in the temperature dependence of the specific heat
\cite{huang1,bpk1}.

Recent theoretical studies on BEC have revealed interesting 
effects of finite number of particles and anisotropic traps
\cite{exp2,dk1,kd1,kt1,gh1,hhr1,mullin1,politzer1,bdl1,ho1,rojas1,yws1,dh1,pathria}.
van Druten and Ketterle \cite{dk1} introduced the 
concept of two-step BEC to explain the observation that  
the cusp in the specific heat appears 
at a temperature $T_{cusp}$ significantly higher than the BEC critical
temperatue $T_c$ in anisotropic harmonic traps.  
This phenomena comes about from the changes in the 
occupations of states in the anisotropic trap with different
characters as the temperature is lowered.  In the present work, we study
the intermediate steps towards BEC in details.  We introduce the concept of
dividing the set of all single particle states into eight subsets, each
of which consisting of states with different characters.  Such a division
is especially useful in treating particles in anisotropic traps in that
the states in the subsets can be classified into those corresponding to
zero-point motion, excited states of oscillators in one, two, and
three dimensions.  The occupations in these subsets can be evaluated
numerically or by invoking a continuum approximation using 
the concept of the density of
states.  The possible multi-step phenomena associated with BEC can then be
described in terms of the changes in the relative populations in the
different subsets as the temperature is lowered.  We found that
pheneomena involving 
more than two steps can occur in a general class of highly
anisotropic traps.  We also found that two-step BEC is possible in
anisotropic traps with $\omega_{1} = \omega_{2} \ll \omega_{3}$, in contrast
to previous claims \cite{dk1}.  The position of the cusp in the specific
heat can also be studied within the context of subsets of states.  The
occupations in the subsets, within the continuum approximation, can be
expressed in terms of Bose functions $g_{n}(z)$.  The Bose functions can be
classified into two classes according to their behaviour at $z=1$.  Our
results indicate that the cusp in the specific heat appears in the vicinity
of the crossovers in the populations between subsets of states involving
Bose functions in {\em different classes} as temperature decreases.   

The plan of the paper is as follows.  The eight subsets of states are
introduced in Sec.2 together with the method of numerical calculations.
Section 3 gives a short discussion on the density of states within the
continuum approximation.  In Sec.4, we present our results demonstrating
the possibility of multi-step phenomena towards BEC in anisotropic harmonic
traps.  Section 5 gives a discussion on the cusp in the specific heat in
terms of the crossover between populations in subsets involving Bose
functions of different classes.  A brief summary is given in Sec.6.

\noindent {\bf 2. Eight subsets of single-particle states}

We consider non-interacting bosons trapped in a three-dimensional (3D) 
anisotropic harmonic traps described by the characteristic frequencies 
$\omega_{1}$, $\omega_{2}$, and $\omega_{3}$ in the $\hat{x}$, $\hat{y}$, 
and $\hat{z}$-axis.  The single particle energy spectrum can
be characterized by
the quantum numbers $n_{1}$, $n_{2}$, and $n_{3}$.  To facilitate our 
discussion, we divide the set  
$S=\{(n_{1},n_{2},n_{3}):n_{1},n_{2},n_{3}=0,1,2,\cdots \}$ of all 
single particle states into eight subsets with different properties, i.e., 
$S=S_0\bigoplus
S_1\bigoplus S_2\bigoplus S_3\bigoplus S_{12} \bigoplus S_{13}\bigoplus
S_{23}\bigoplus S_{123}$. Here $S_{0}=\{(0,0,0) \}$ consists only of the
overall ground state. The subsets $S_{1}=\{ (n_{1},0,0):n_{1}=1,2,\cdots \}$
, $S_{2}=\{ (0,n_{2},0):n_{2}=1,2,\cdots \}$, and $S_{3}=\{
(0,0,n_{3}):n_{3}=1,2,\cdots \}$ consist of states in which 
the lowest energy state of the corresponding  
harmonic oscillators in two of the three spatial directions
are occupied and the excited states in only one spatial
dimension are occupied.  
Hence, $S_{1}$, $S_{2}$, and $S_{3}$ include states which are 
one dimensional (1D) in character.  
Similarly, the subsets $S_{12}$, $S_{13}$, and $S_{23}$ consist
of states which are two dimensional (2D) in character in that 
they consist of 
states in which one and only one of the three quantum numbers vanishes,
and hence include states which are excited 
in the $xy$, $xz$, and $yz$ planes, respectively. The subset $
S_{123}=\{ (n_{1},n_{2},n_{3}): n_{1},n_{2},n_{3}=1,2,\cdots \}$ consists of
all the 3D excited states with nonzero $n_{1}$, $n_{2}$, and $n_{3}$. 

The mean occupation number $n_p$ in a
single-particle state with energy $\epsilon_p$ is given, within the
grand-canonical ensemble, by 
\begin{equation}
n_{p}=\frac{1}{e^{\beta(\epsilon_{p}-\mu)}-1},  \label{eq:population}
\end{equation}
where $\beta\equiv 1/k_{B}T$ is the inverse temperature with $k_{B}$ being
the Boltzmann constant. The chemical potential $\mu$ or the fugacity $
z\equiv \exp(\beta\mu)$ is determined by the mean number of particles $N$ in
the system via $N=\sum_{p}n_{p}$. Defining \cite{dh1} the factor 
$
B_{j}\equiv\sum_{p}\exp(-j\beta\epsilon_{p}),
$
where the summation is over all the single-particle states labelled by $p$,
then we have 
\begin{equation}
N=\sum_{j=1}^{\infty} z^{j}B_{j}.  \label{eq:number}
\end{equation}
Note that all the effects of the confining potential on the thermodynamic
properties are included in the factor $B_{j}$ through the energy spectrum $
\epsilon_{p}$.
For the 3D parabolic potential with possibly different frequencies in
different directions, the single-particle eigen-energies are given by 
$
\epsilon_{n_{i}}=\sum_{i=1}^{3}n_{i}\hbar\omega_{i},
$
where $\omega_{1}\le \omega_{2}\le \omega_{3}$ are the frequencies
characterizing the confinement in different directions.
The quantum numbers $n_1$, $n_2$, and $n_3$ take on non-negative integers,
and the zero-point energy has been absorbed into the definition of the zero
of energy. 
Once the fugacity $z$ is determined by Eq.(\ref{eq:number}),
the internal energy $U$
can be calculated directly via 
$
U=-\sum_{j=1}^{\infty}z^{j}(\partial B_{j}/\partial \beta)/j,
$
and the heat capacity $C_v= (\partial U/\partial T)_{N,v}$ can also be obtained. The
subscript $v$ in $C_{v}$ indicates that this quantity is analogous to the
heat capacity at fixed volume in the sense that the oscillators' frequencies
are held fixed.  The occupation in each of the subsets of states can also 
be evaluated as a function of temperature.  Equation (2) forms the basis of 
our discussion of multi-step phenomena associated with BEC.

\noindent {\bf 3. Continuum approximation using DOS}

In addition to studying the mean occupation in each subset of states 
numerically, we can also study the occupations analytically by invoking the 
notion of the density of states (DOS) 
\cite{huang1,bpk1,dk1,kd1,kt1,gh1,hhr1,mullin1}.  The DOS for each of the
subsets can readily be written down.  However, one should keep in mind 
the fact that the different subsets of states have {\em different} 
lower limits of energy when a sum over states is turned into an integral, 
a point that has become a source of confusion in the literature 
\cite{bpk1,kd1}.  
The mean
occupation numbers in the eight subsets of states can then be 
expressed as: 
\begin{equation}
\left \{ 
\begin{array}{llll}
N_{0} & \approx & g_{0}(z),& \\ 
N_{i} & \approx & \frac{1}{\Omega_{i}} g_{1}(ze^{-\Omega_{i}^{*}}),& i=1,2,3, \\
N_{i\ell} & \approx & \frac{1}{\Omega_{i}\Omega_{\ell}} g_{2}(ze^{-\Omega_{
\ell}^{*}}),&\ell=12,13,23, \\ 
N_{123} & \approx & \frac{1} {\Omega_{1}\Omega_{2}\Omega_{3}} g_{3}(ze^{-
\Omega_{123}^{*}}).&
\end{array}
\right.  
\label{eq:integral}
\end{equation}
Here $\Omega_{i}\equiv \hbar\omega_{i}/k_{B}T \ (i=1,2,3)$, 
$\Omega^{*}\equiv \epsilon^{*}/k_{B}T$, and  
the proper lower limits of energy for integrations involving the 
DOS for subsets corresponding to excited states with 1D, 2D, and 3D 
characters should
be taken to be $\epsilon_{i}^{*}=\hbar\omega_{i}/2$ ($i=1,2,3$), $
\epsilon_{i\ell}^{*}=\hbar(\omega_{i}+\omega_{\ell})/2$
($i\ell=12, 13, 23$),
and $\epsilon_{123}^{*}=\hbar(\omega_{1}+\omega_{2}+\omega_{3}+ \sqrt{
\omega_{1}^2+\omega_{2}^2+\omega_{3}^2} )/2$, respectively. 
The Bose functions 
$g_{n}(z)$ are 
defined\cite{huang1} by $g_{n}(z)=\sum_{j=1}^{\infty}z^j/j^n$.
These functions can be divided into {\em two classes}.
The first class consists of  
Bose functions with $n >1$, and they are finite at $z=1$.  
The second class consists of Bose functions with $n\le 1$,  
and they  diverge at $z=1$.  For example, 
$g_{0}(z)=z/(1-z)$ and $g_{1}(z)=-\ln(1-z)$ diverge at $z=1$.  
The results in Eq.(3) also give us insight into the multi-step phenomena 
associated with BEC as 
the dominant subset of states carrying a 
substantial fraction of the total particles changes from one subset to 
another as the temperature is lowered. 
Similar approximate expressions for the DOS can be obtained in various different ways
such as the 
Euler-Maclaurin summation\cite{kt1}, the zeta function\cite{hhr1,shiokawa1}, or the simple 
series\cite{dk1}. In the approximation scheme of Ref.\cite{dk1}, the single-particle
states are divided into four groups.

\noindent {\bf 4. Multi-step BEC}

BEC is usually described as the onset of
a macroscopic occupation of the overall
ground state, i.e., the subset $S_{0}$, at the critical temperature $T_{c}$. 
Very often, it is accompanied by the presence of a cusp in the specific 
heat at the same temperature.  
In recent experiments\cite{exp1,exp2}, the direct evidence for 
BEC is taken to be the abrupt increase
in the peak density at the center of the trapped atomic cloud. 
Novel experimental techniques have made possible the measurements in the 
spatial and velocity distributions of the trapped bosons.  
If the subset of states
containing the largest fraction of particles changes from
one subset to another consisting of states with a lower dimensional character 
as temperature decreases, the
changes in the spatial and velocity distributions
should also be measurable.  In fact, these changes in the characteristic 
of the trapped bosons have been taken to be the 
indication of 
two-step phenomena in BEC \cite{dk1,kd1}.  Here we will study the possible 
multi-step phenomena in BEC due to the changes in the occupations in the 
subsets of states as temperature decreases\cite{shiokawa1}.  

Following Ref.[6], a ``saturation" temperature can be defined for each 
subset of excited states.  Consider a hypothetical system consisting only 
of one subset of states, say $S_{123}$, plus the overall ground state.  At 
the saturation temperature $T_{3D}^{(0)}$, the population in the 
set of excited states saturates. 
From Eq.(\ref{eq:integral}), $T_{3D}^{(0)}$ is approximately given by 
$T_{3D}^{(0)}=(0.94\hbar/k_B)(N\omega_1\omega_2\omega_3)^{1/3}$. 
Similarly, the ``saturation" temperature for the subsets $S_{ij}$ ($ij=
12,13,23$) with 2D character is given by 
$T_{2D}^{(0)}=(0.78\hbar/k_B)(N\omega_i\omega_j)^{1/2}$; 
and that for the subsets $S_{i}$ ($i = 1,2,3$) of 1D character is given by 
$T_{1D}^{(0)}\approx (N\hbar\omega_i/k_B)\ln(2N)$ 
\cite{dk1,kd1}. 
For example, only the subsets $S_{123}$ and $S_{0}$ are responsible for
the BEC in 3D {\em isotropic} harmonic 
trap, and the ``saturate temperature'' 
$T_{3D}^{(0)}=0.94N^{1/3}\hbar\omega/k_B$ 
is the corresponding critical temperature below which 
the overall ground state becomes macroscopically occupied. 

For suitably chosen $\omega_{1}$, $\omega_{2}$, and $\omega_{3}$, 
the competitions among the different subsets may lead to complicated 
and interesting phenomena.  
Consider $\omega_3:\omega_2:\omega_1=2.1\times 10^4:4.4\times 10^3:1$. 
This choice of parameters corresponds to $T_{3D}^{(0)}=1.21T_{2D}^{(0)}$
and $T_{2D}^{(0)}=2T_{1D}^{(0)}$.  Figure 1 shows the typical behavior 
of three-step phenomena associated with BEC for $N=10^5$ particles in such 
anisotropic traps.  
The temperature is expressed in reduced temperature defined by 
$\tau\equiv k_BT/(\hbar^3\omega_1\omega_2\omega_3)^{1/3}$. 
As the temperature is lowered, the subset containing  
the largest fraction of particles
changes from $S_{123}$ to $S_{12}$ at about $\tau_1=1.5$ (the first step), 
then from $S_{12}$ to $S_{1}$ at $\tau_2=0.6$ (the second step), 
and finally to $S_{0}$  
at about $\tau_3=0.2$ (the third step).
On the other hand, the BEC critical temperature at which $N_{0}$ 
becomes appreciable is at about $\tau_c=0.5$.  Figure 1 also shows  
the behavior of the heat capacity.  The cusp in the heat capacity appears  
at $\tau_{cusp}=0.65$, which is higher than $\tau_{c}$.  

Figure 2 shows the results for $N=10^5$ bosons in an anisotropic trap with
$\omega_{1} = \omega_{2} = \omega_{3}/500$.  It was reported in 
Ref.\cite{dk1} that two-step phenomena will not be observed in such system
since $\tau_{cusp}$ and $\tau_{c}$ are nearly the same ($\tau_{cusp} 
\approx \tau_{c} = 0.65$).  The choice of parameters corresponds to 
the case in which 
$\omega_1=\omega_2\ll \omega_3$, and 
we have 
$T_{3D}^{(0)}/T_{2D}^{(0)}=1.21(\omega_3^2/\omega_1^2N)^{1/6}\approx 1.41$.
However, our results demonstrated 
that the particles start to condense from the subset  
$S_{123}$ into $S_{12}$ at a temperature $\tau_1\approx 2.0$, and 
then to $S_{0}$ at a lower temperature.  Such two-step   
phenomena should be observable.

\noindent{\bf 5. Cusp in specific heat and crossover between subsets}

Our approach of focusing on the occupation in the subsets of states also 
shed light on the position of the cusp in the heat capacity.  In the 
standard textbook case of ideal Bose gas in the thermodynamical limit, 
the temperature $T_{cusp}$ corresponding to the position of the cusp 
coincides with the critical temperature $T_{c}$ of BEC.  However, there 
exists in general no such restrictions requiring the coincidence of the 
two temperatures for systems with finite number of particles in anisotropic
traps \cite{dk1}.  Recall that the Bose functions $g_{n}(z)$ can be 
classified into two classes depending on their behavior at $z=1$.  For 
harmonic traps, the occupations in $S_{123}$, $S_{12}$, $S_{13}$, $S_{23}$ 
involve $g_{n}(z)$ with $n>1$ (the first class) while the occupations 
in the subsets $S_{0}$, $S_{1}$, $S_{2}$, $S_{3}$ involve $g_{n}(z)$ with 
$n \leq 1$ (the second class).  An interesting question is that whether
there will exist 
two or more cusps in the specific heat in multi-step phenomena
associated with BEC.  We have studied the temperature dependence of the
specific heat for a wide range of ratios of $\omega_{1}$, $\omega_{2}$, and
$\omega_{3}$.  Our results indicate that there is at most
one peak in the temperature dependence of the heat capacity, with the peak 
being rounded off due to the finite number of particles in our calculations. 
The position $T_{cusp}$ of the cusp appears to be the results of a 
combination of crossovers in the occupations between subsets involving
$g_{n}(z)$ belonging to the two {\em different classes}.  
The crossovers at which
the particles condense from one subset to another involving Bose functions
belonging to the same class do not lead to a cusp in the specific heat. 
For the system 
considered in Fig.1, one such crossover occurs between the subsets 
$S_{123}$ and $S_{1}$ at $\tau \approx 0.85$, which coincides with a 
rapid increase in the heat capacity as temperature decreases.  Another 
crossover between the subsets $S_{12}$ and $S_{1}$ at $\tau = 0.65$ 
coincides with the cusp in the heat capacity.  For the case in Fig.2,
the crossover between the two
classes of Bose functions occurs between the subsets $S_{123}$ and $S_{0}$
at $\tau \sim 0.65$, which coincides with the position of the cusp in the 
heat capacity.  We further re-examine the case of $10^{6}$ bosons in a 
highly anisotropic trap with $\omega_{2} = \omega_{3} = 5.6\times 10^{6} 
\omega_{1}$ corresponding to $T_{3D}^{(0)} = 2 T_{2D}^{(0)}$.  This 
system was studied in Ref.\cite{dk1} to demonstrate the two-step 
phenomena.  Results are shown in Fig.3.  The critical temperature 
$T_{c}$ is appreciably lower than $T_{cusp}$.  The position of the cusp 
coincides with the range of temperatures in which the two crossovers, one 
between ($S_{12},S_{13}$) and $S_{1}$, and another between $S_{123}$ and 
$S_{1}$, involving different classes of Bose functions occur.  

It should be pointed out that our analysis reproduces
the standard results in isotropic traps.  For
isotropic harmonic traps in 3D (2D),
BEC is accompanied by a crossover in occupations
between $S_{123}$ ($S_{12}$) and $S_{0}$ and hence $T_{cusp} = T_{c}$ 
as $S_{123}$ ($S_{12}$) and $S_{0}$ involve different classes of $g_{n}(z)$. 
For 1D harmonic traps, BEC is accompanied by the crossover between $S_{0}$ 
and $S_{1}$, both of which involve Bose functions of the same class and 
hence there exists no cusp in the heat capacity.  

Extending our argument to bosons in 3D boxes with dimensions 
$L_{1}$, $L_{2}$ and $L_{3}$ in different directions is interesting\cite{dk1}.  The
occupation in a subset of states with $d$-dimensional character involves
the Bose function $g_{d/2}$ \cite{huang1}.
Hence only the subset $S_{123}$ including
excited states in all three Cartesian coordinates involves $g_{n}(z)$ with 
$n=3/2 >1$.  All the other subsets consist of 
states of lower-dimensional characters
and involve $g_{n}(z)$ with $n \leq 1$.
Thus, BEC occurs as long as the spatial dimensions of the box
and the number of particles allow the subset $S_{123}$ to be occupied.  
For a box with $L_{1} \ll L_{2} \ll L_{3}$, we expect a three-step
BEC to occur with the largest occupation changes, as temperature decreases,
from $S_{123}$ to $S_{23}$,
then to $S_{3}$, and finally to $S_{0}$ with the cusp of the specific 
heat appearing in the vicinity of the crossover between $S_{123}$ 
and $S_{23}$.   The crossover from $S_{23}$ to $S_{3}$ and
from $S_{3}$ to $S_{0}$ do not lead to a cusp in the specific heat since
the subsets involve Bose functions belonging to the same class.
For a box with $L_{1} \ll L_{2} = L_{3}$, we expect
a two-step phenomena with the cusp of the specific heat appearing at a 
temperatue near the crossover between the subsets $S_{123}$ and
$S_{23}$ which is higher than the BEC critical temperature.  For a box
with $L_{1} = L_{2} \ll L_{3}$, we expect that the cusp to appear  when
near the crossover between $S_{123}$ and $S_{3}$ at a temperature higher
than the BEC temperature.

\noindent{\bf 6. Summary}

In summary, we have studied the intermediate steps towards BEC for finite
number of bosons confined by anisotropic harmonic traps.  Using the concept
of dividing the set of all single particle states into subsets consisting of
states with different characters, two-step and multi-step phenomena
associated with BEC can be descibed in terms of the crossovers in the
populations of these subsets as temperature is lowered.  Cases with
three-step processes are demonstrated.  The cusp of the specific heat
is found to be associated with the crossover between subsets involving
Bose functions of different classes.  The general method adopted in the
present work can readily be extended to study bosons in other types of
traps.  

\noindent {\bf Acknowledgments}

The author thanks Prof. P. M. Hui for useful discussions and a critical reading
of the manuscript.
He also would like to thank Prof. B.-L. Hu for drawing his attention to Ref.[19]. 
This work was  supported by national ``Climbing Project'' in theoretical physics,
and partly by the GuangDong Provincial Natural Science Foundation of China.

\begin{figure}
\caption{
Numerical results for $N=10^5$ particles in a 
3D anisotropic harmonic trap with 
$\omega_3:\omega_2:\omega_1=2.1\times 10^4:4.4\times 10^3:1$. 
The solid lines show the temperature dependence of the populations
$N_0/N$ in the subset $S_{0}$,  $N_1/N$, $N_{12}/N$,
and $N_{123}/N$ in the subsets $S_{1}$, $S_{12}$ and $S_{123}$, 
respectively.  The populations
in the other subsets are too small to be shown. 
The line with circles gives 
the rescaled specific heat as a function of
the reduced temperature 
$\tau \equiv k_{B}T/(\hbar^{3}\omega_{1}\omega_{2}\omega_{3})^{1/3}$.}
\label{Fig.1}
\end{figure}
\begin{figure} 
\caption{Two-step BEC occurs in a Bose gas of $N=10^5$ particles in 
an anisotropic trap with $\omega_{3}:\omega_{2}:\omega_{1}=500:1:1$. 
The particles condense  
from the subset $S_{123}$ into the subset $S_{12}$ 
at a temperature
significantly higher than the BEC critical temperature, 
showing features of a two-step BEC.  However, the temperature 
for specifc heat cusp
is close to the BEC critical temperature.
}
\label{Fig.2}
\end{figure}
\begin{figure} 
\caption{Typical behaviour of two-step BEC of $N=10^6$ particles
in a highly 
anisotropic harmonic trap with $\omega_3=\omega_2=5.6\times 10^6\omega_1$, 
corresponding
to $T_{3D}^{(0)}=2T_{1D}^{(0)}$. The condensations from the subsets 
$S_{123}$, $S_{12}$, and $S_{13}$ 
to $S_{1}$ are responsible for the cusp in the specifc heat.}
\label{Fig.3}
\end{figure}
\begin{figure}[btp]
\begin{center}
\leavevmode
\epsfbox{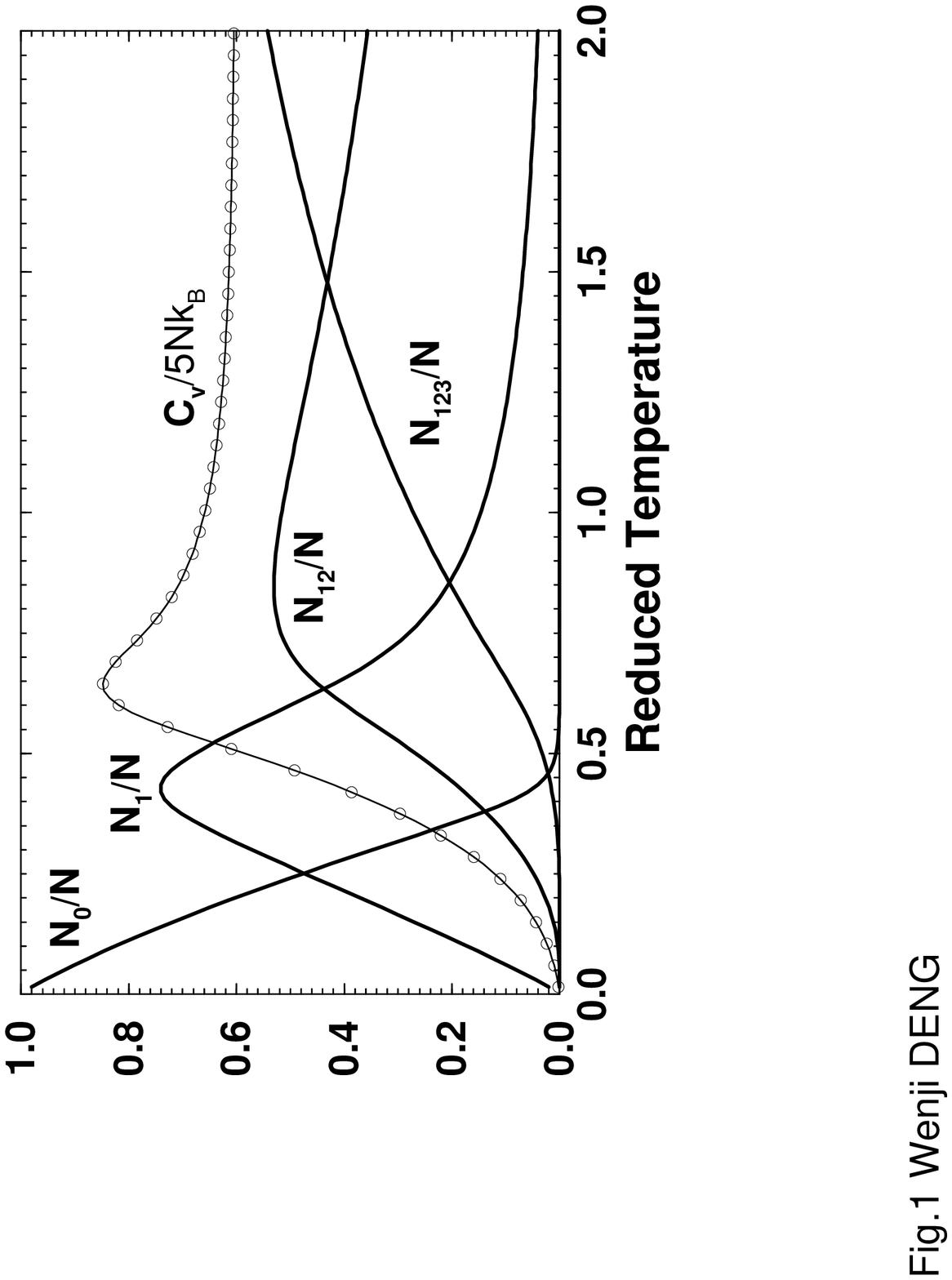}
\end{center}
\end{figure}
\begin{figure}[btp]
\begin{center}
\leavevmode
\epsfbox{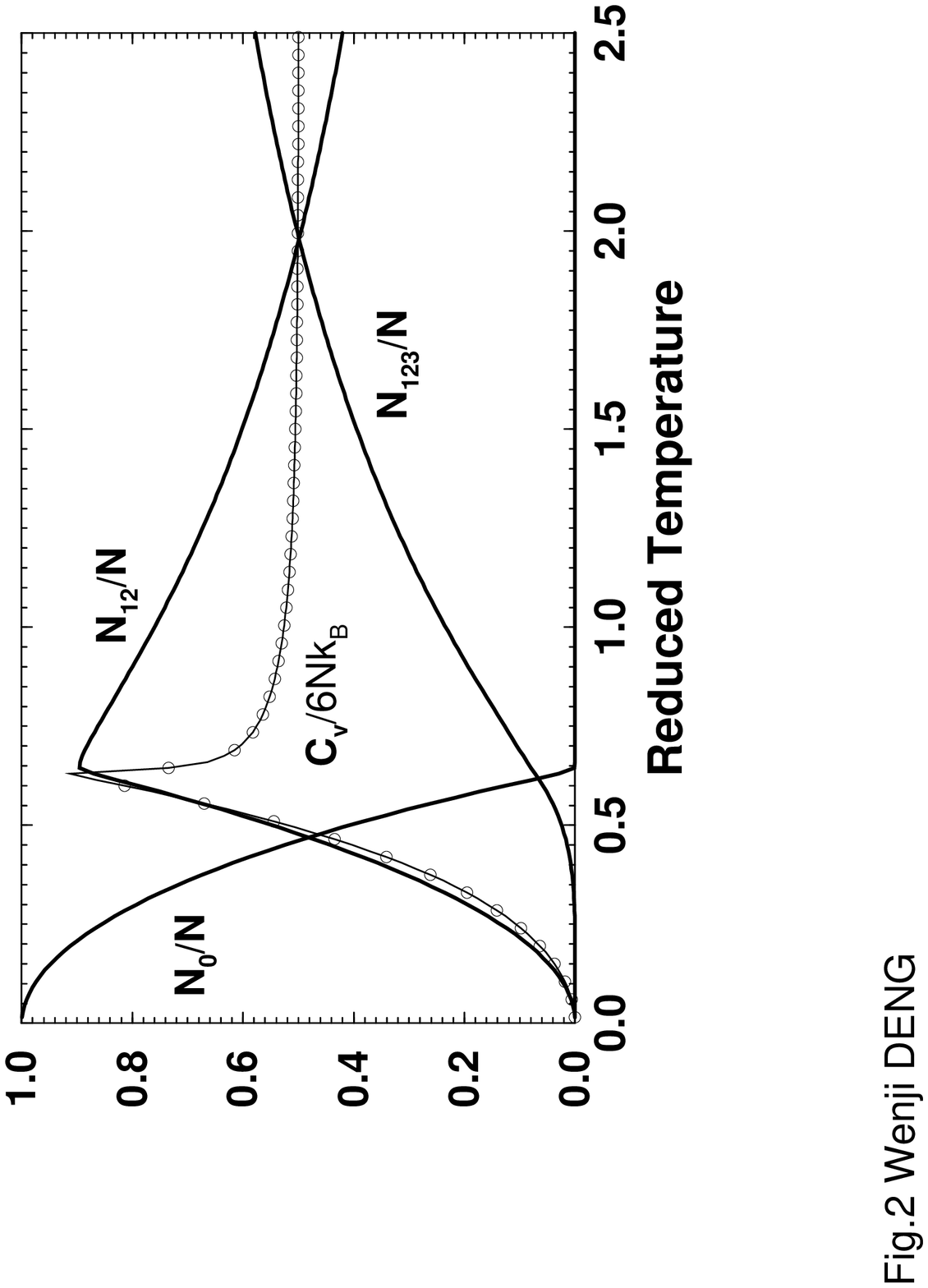}
\end{center}
\end{figure}

\begin{figure}[btp]
\begin{center}
\leavevmode
\epsfbox{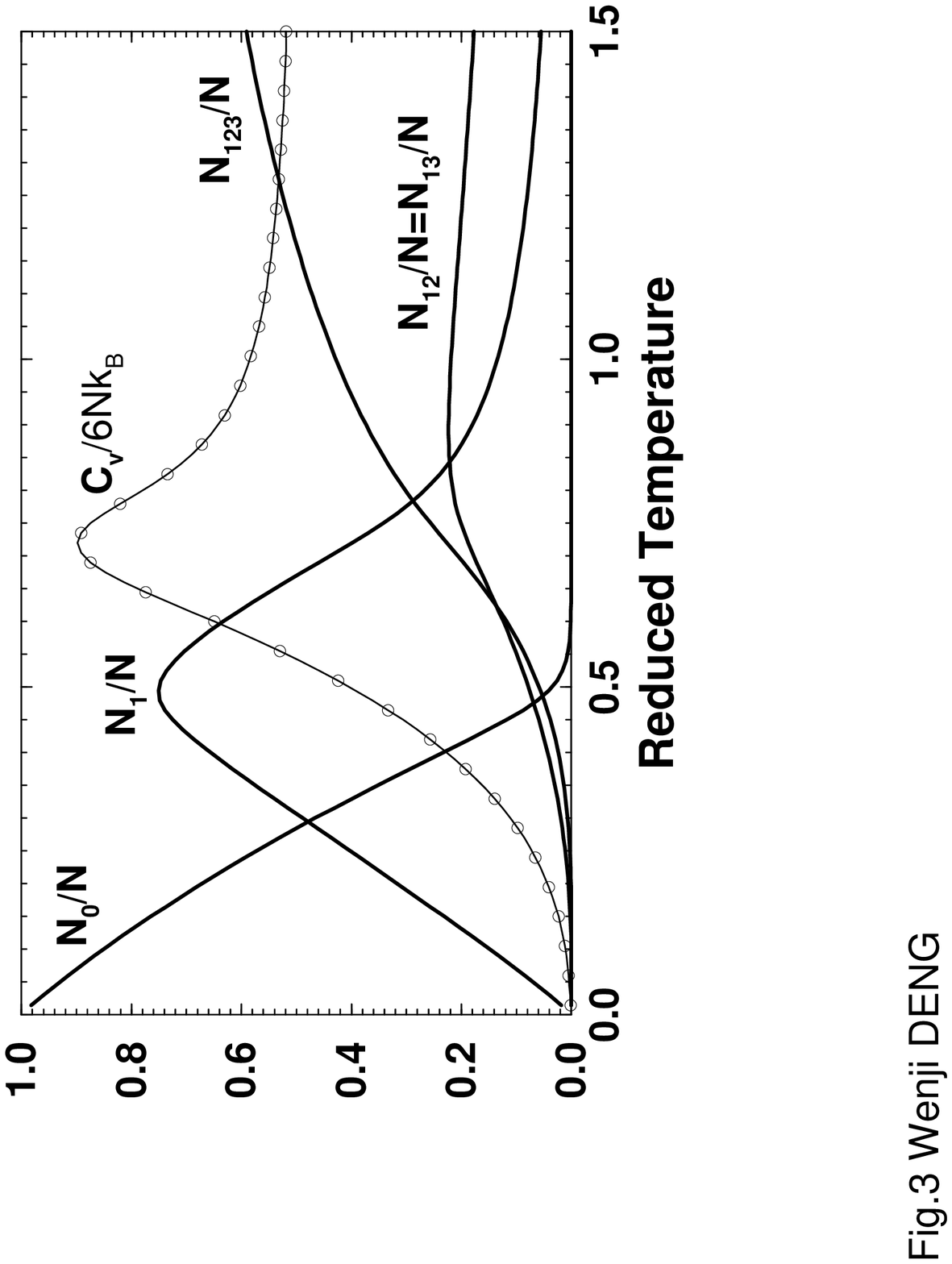}
\end{center}
\end{figure}

\end{document}